\documentclass[aps,prd,twocolumn,showpacs]{revtex4}
\usepackage{graphicx}
\begin{document}
\title{Evolution of shower parton distributions in a jet from quark recombination model}
\author{Z.G. Tan and C.B. Yang}
\affiliation{Institute of Particle Physics, Hua-Zhong Normal University,
Wuhan 430079, P.R. China}
\date{}
\begin{abstract}
The evolution of shower parton distributions in a jet is investigated in the
framework of quark recombination model. The distributions are parameterized
and the $Q^2$ dependence of the parameters is given by polynomials of $\ln Q^2$
for a wide range of $Q^2$.

\end{abstract}

\maketitle
\section{Introduction}
The study of hadron production in high energy collisions is the only way
to understanding structure of hadrons and the dynamics of the collision process.
Because of asymptotic freedom at short distance and quark confinement at long distance
of the strong interaction, hadron formation is a soft process which cannot be calculated
perturbatively from the first-principle for the strong interactions--the quantum
chromodynamics (QCD). Therefore some model considerations are needed whenever hadron
formation is involved in a process. One of the most frequently used models for hadron
production is the parton fragmentation model. This model was first proposed in studying
the hadron production in $e^+e^-$ annihilation process  \cite{frag}. In such a process,
$e^+e^-$ annihilate into a virtual photon which then materializes into a $q\bar{q}$ pair.
The produced $q$ and $\bar{q}$ move away from each other and each fragments or cascades
into a collection of hadrons (or jet). Fragmentation functions have been used to
parameterize the non-perturbative aspects of jets from initiating quarks and gluons
in last a few decades. As stated clearly in \cite{app}, however, this approach ``is
not meant to be a theory. It is simply a parameterization that incorporates many of
the expected features of fragmentation.'' The fragmentation functions does not tell
us how the hard partons fragment into jets. If hard parton fragmentation were the only
way for the production of hadrons with large transverse momentum, the ratio between
two hadron species, for any given transverse momentum, would be a process independent
constant dictated by the fragmentation functions. However, experimental data
at the Relativistic Heavy Ion Collider (RHIC) at Brookhaven have shown that the proton
to pion ratio can be about 1 for $p_T\sim 3$ GeV/$c$ in central AuAu collisions at
$\sqrt{s_{NN}}=200$ GeV \cite{propion}, much larger than the corresponding value in
$pp$ collisions. This experimental fact indicates that the study of the mechanism
of parton fragmentation is necessary in order to understand the colliding system
dependence of hadron production in a jet.

A theoretical microscopic description of parton fragmentation is suggested in \cite{hy1}
based on quark recombination approach \cite{recom1,recom2} for hadron production. The
physics issue in the microscopic description of fragmentation is that every quark
and gluon with high virtuality is assumed to evolve by emanating quark-antiquark pairs
and gluons with lower virtualities into a shower of partons with low transverse momentum
relative to the initiating parton direction. The shower partons recombine then to form
the final state hadrons. In this way, the fragmentation functions can be related to
the shower parton distributions in a jet. In high energy hadronic and nuclear
collisions, there exist some soft partons with
low virtuality and some hard partons produced in hard collisions with high momentum
transfer. Those hard partons may have very high virtuality and, according to the
principles of QCD, will evolve into parton showers. Those
shower partons, together with the soft partons, then recombine into final
state hadrons detected in experiments, as described in the recombination model.

In \cite{hy1} shower parton distributions initiated by a hard parton
with virtuality $Q^2=100$ $({\rm GeV}/c)^2$ were obtained from the fragmentation
functions determined by fitting experimental data for high energy $e^+e^-$
annihilation and hadronic collisions. Once the shower parton distributions are obtained,
it is natural to apply them in modeling hadron production in other collisions.
In \cite{recom2} it is shown that the recombination of soft partons and shower
partons is important in reproducing the transverse momentum spectra of produced
particles in both AuAu and dAu collisions and in explaining the centrality dependence
of the Cronin effect \cite{cronin} in dAu collisions.

In this paper, we investigate the parton distributions in a jet for $Q^2$ from 2 to 1000
$({\rm GeV}/c)^2$. We also parameterize the $Q^2$ dependence of the parameters for
the parton distributions for easy use in the recombination model description for
hadron productions in different collision processes. The application of the parton
distributions will be left for later work. This paper is organized as follows. In section
II, we briefly discuss the recombination model description of parton fragmentation. Then
in section III, we discuss the $Q^2$ evolution of the parton distributions. The last
section is for a short discussion.

\section{Shower distributions in a jet from recombination model} For brevity,
one can introduce following notation for any two parton distributions
$f_1(x)$ and $f_2(x)$ in a jet,
\begin{eqnarray*}
\{f_1f_2\}(x_1,x_2)\equiv & 0.5\left(f_1(x_1)f_2\left({x_2\over 1-x_1}\right)+\right.\\
& \left.f_1\left({x_1\over 1-x_2}\right)f_2(x_2)\right)\ .
\end{eqnarray*}
Here $x_1, x_2$ are momentum fractions of two partons relative to the initiating
parton. In this expression, kinetic constraint is imposed so that $x_1+x_2<1$
for any pair of two quarks.

In the quark recombination model for fragmentation the generic formula for the production
of a meson composed of a quark $q$ and an antiquark $\bar{q}^\prime$ is \cite{rh,hy1}
\begin{eqnarray}
xD(x) = \int^x_0 {dx_1 \over x_1} \int^x_0 {dx_2 \over
x_2}\{q\bar{q}^\prime\}(x_1, x_2) R (x_1, x_2, x) \ ,
\label{1}
\end{eqnarray}
where $q(x_1)$ and $\bar{q}^\prime(x_2)$ are the parton distributions of a quark
$q$ at momentum fraction $x_1$ and an antiquark $\bar{q}'$ at
$x_2$ in a jet, and $R (x_1, x_2, x)$ is the recombination function (RF) for the
formation of a meson at $x$ from the quark $q$ and antiquark $\bar{q}^\prime$.
In the recombination model, the gluons are assumed to convert first to
quarks and antiquarks before hadron formation. So in last formula no gluon
distribution is considered.

\begin{figure}[tbph]
\includegraphics[width=0.45\textwidth]{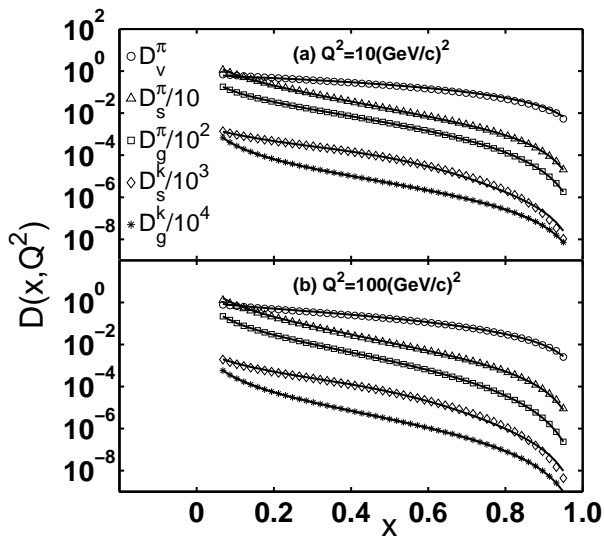}
\caption{Fitted fragmentation functions for valence quark to pion, sea quark to pion,
gluon to pion, and sea quark to kaon, gluon to kaon, at (a) $Q^2=10$ and (b) 100 $({\rm
GeV}/c)^2$. The fitted results are shown as solid curves, and the parameterizations from
KKP (for gluon jets) and Kretzer (for quark jets) by different symbols for different
fragmentation functions.}
\end{figure}

At this stage, we only consider production of hadrons composed of light quarks
($u, \bar{u}, d, \bar{d}, s, \bar{s}$) fragmented from light hard partons $(u, \bar{u}, d,
\bar{d}, s, \bar{s}, g)$. If we assume SU(2) symmetry and symmetric $u\bar{u}$ and
$d\bar{d}$ sea quark distributions, there are five independent parton distributions.
Three of them are for parton distributions in a jet initiated by light quarks. They are
the non-singlet distribution $K_{\rm NS}$ of valence quark with the same flavor as the
initiating (anti)quark, the non-strange sea quark distribution $L$, the strange sea
quark distribution $L_s$. The other two are for the non-strange light quark distribution
$G$  and strange quark distribution $G_s$ in a gluon jet.
Five fragmentation functions are needed in order to determine those five parton
distributions. We work with fragmentation functions for the pion production $D^{\pi}_S$
from sea quarks, $D^{\pi}_V$ from valence quark, $D^{\pi}_G$ from gluon jet, and
for K production $D^{K}_S$ from sea quarks, $D^{K}_G$ from gluon jet. The corresponding
expressions for the fragmentation functions in terms of the parton distributions
are given in the quark recombination model as \cite{hy1}
\begin{equation}
xD^{\pi}_S(x)=\int {dx_1\over x_1} {dx_2\over
x_2}\{LL\}(x_1,x_2)R^{\pi}(x_1, x_2, x) ,
\label{2}
\end{equation}
\begin{equation}
xD^{\pi}_V(x)=\int {dx_1 \over x_1} {dx_2 \over
x_2}\{KL\}(x_1,x_2)R^{\pi} (x_1, x_2, x),
\label{3}
\end{equation}
\begin{equation}
xD^{\pi}_G(x)=\int {dx_1 \over x_1} {dx_2 \over
x_2}\{GG\}(x_1,x_2)R^{\pi} (x_1, x_2, x) ,
\label{4}
\end{equation}
\begin{equation}
xD^K_S(x)=\int {dx_1 \over x_1} {dx_2 \over x_2}
\{LL_s\}(x_1,x_2)R^K (x_1, x_2, x) ,
\label{5}
\end{equation}
\begin{equation}
xD^K_G(x)=\int {dx_1 \over x_1} {dx_2 \over x_2}
\{GG_s\}(x_1,x_2)R^K (x_1, x_2, x) .
\label{6}
\end{equation}
In last equations $R^\pi$ and $R^K$ are RF for pions and kaons, respectively.
Here, both the fragmentation functions and the parton distributions in a jet
depend on the virtuality $Q^2$ of the initiating parton.

Before one can start determining the parton distributions from above equations,
one needs to make
a decision on choosing parameterizations of the fragmentation functions given by different
groups. In fact, the determination of fragmentation functions is not as well as
supposed to be.  Different physics considerations are input in different parameterizations
of the fragmentation functions, and sometimes the obtained results are not
consistent with each other. The most frequently used parameterizations of the
fragmentation functions are given by Kniehl, Kramer, P\"otter (KKP) \cite{kkp,kkp2}
and by Kretzer \cite{sk}. In \cite{sk} it is assumed that valence quark dominates
the production of hadrons with large momentum fraction and sea quarks contribute
more to hadron production at small momentum fraction. In KKP parameterizations, no such
physics consideration is input, but they used more recent experimental results involving
gluons. In our paper, we adopt Kretzer parameterization of the fragmentation functions
for jets initiated by quarks or antiquarks for a better consideration of valence quark
contributions, but use KKP parameterization for gluon jets.

In fitting the fragmentation functions from quark recombination model, as given in Eqs.
(\ref{2}-\ref{6}), we assume the parton distributions to be the form
\begin{equation}
f(z)=Az^a(1-z)^b(1+cz^d)\ ,
\end{equation}
with five parameters $A, a, b, c$ and $d$, and $z$ the momentum fraction of the quark
relative to the initiating parton. For the valence quark distribution,
a condition is imposed
\begin{equation}
\int {dz\over z} K_{\rm NS}(z)=1\ ,
\end{equation}
because the total number of valence quark in a jet cannot be more than one.

The fragmentation functions for pion and kaon have been fixed in \cite{valon} as
\begin{eqnarray}
R^\pi(x_1,x_2,x)&=&{x_1x_2\over x^2}\delta\left({x_1+x_2\over x}-1\right)\ ,\\
R^K(x_1,x_2,x)&=& 12{x_1^2x_2^3\over x^5}\delta\left({x_1+x_2\over x}-1\right)\ .
\end{eqnarray}
In the recombination function $R^K$ for kaons, $x_2$ refers to the momentum fraction of
the strange (anti)quark and $x_1$ to that of the non-strange (anti)quark.

Then one can fit the fragmentation functions and get the parton distributions
in a jet for any fixed $Q^2$. The obtained parameters $A, a, b, c$ and $d$ for
the five parton distributions depend on the scale $Q^2$. For $Q^2=10$ and 100
$({\rm GeV}/c)^2$ the fitted results are shown in Fig. 1 as solid curves,
in comparison with the KKP and Kretzer parameterizations. Obviously, the recombination
model can fit the fragmentation functions very well. It should noted that the
fitted curve for $K_{\rm NS}$ agrees with Kretzer parameterization quite well,
in sharp difference from the fitted result in \cite{hy1}, there the parameterization
for the valence quark fragmentation had a significant contribution at small
momentum fraction region, while in Kretzer parameterization the valence
quark fragmentation is assumed to dominate in large momentum fraction region
and no singularity at very small momentum fraction.

The fitted shower parton distribution functions ($pdf(z)$)are shown in Fig. 2
for $Q^2=10$ and 100 $({\rm GeV}/c)^2$. As expected, the non-singlet distribution
$K_{\rm NS}$ for valence quark is hardest among the five distributions. Other
parton distributions show a strong
increase at small $z$ as $z$ goes to smaller values.
\begin{figure}[tbph]
\includegraphics[width=0.45\textwidth]{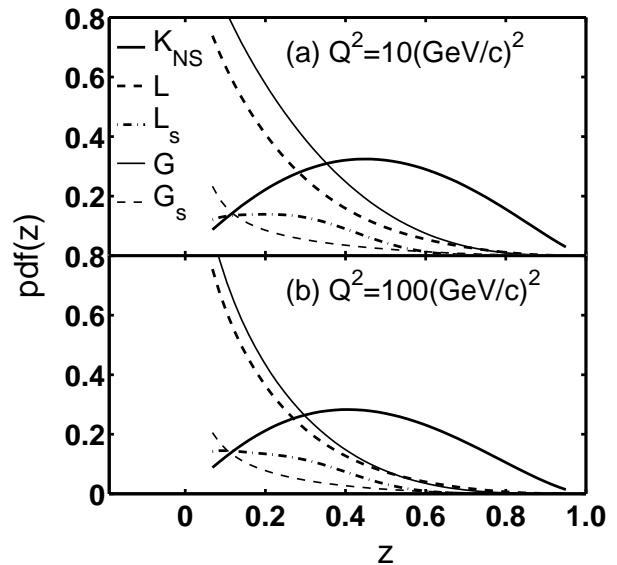}
\caption{Fitted shower parton distribution functions (pdf(z))
in a jet at (a) $Q^2=10$ and (b) 100 $({\rm GeV}/c)^2$.}
\end{figure}

\section{Evolution of parton distributions}
In \cite{hy1} only the fragmentation
functions at fixed $Q^2=100$ $({\rm GeV}/c)^2$ are employed. The obtained
parton distributions are then used to calculate the contributions from both
soft-shower and shower-shower parton recombination to the hadron $p_T$ spectra.
For simplicity, all hard partons are assumed to have the same virtuality $Q^2=100$
$({\rm GeV}/c)^2$ in all applications. In more realistic cases, partons with
different momenta may have different virtualities. This difference in virtuality
may influence hadron productions
from the fragmentation of hard partons and from other quark recombination processes.

\begin{figure}[tbph]
\includegraphics[width=0.45\textwidth]{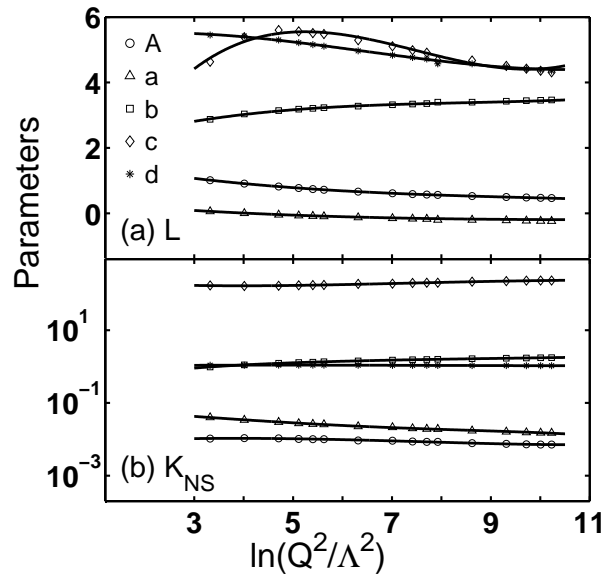}
\caption{$Q^2$ dependence of five parameters for parton distributions $K_{\rm NS}$
and $L$. Solid curves are fitted results by Eq. (\ref{q2}). The QCD scale
$\Lambda$ is chosen to be 0.19 $({\rm GeV}/c)^2$. }
\end{figure}

From perturbative QCD we know that the $Q^2$ evolution of the fragmentation functions
is described by the DGLAP equations \cite{dglap}. From such equations
one can determine the fragmentation
functions at any $Q^2$ if one knows them at one given scale $Q_0^2$. Up to now however,
we did not know the $Q^2$ dependence of parton distributions in a jet initiated by
a hard parton, nor have we had an equation to describe their evolution in a model
independent way. But we can get such a dependence by using Eqs.
(\ref{2}-\ref{6}) from the
$Q^2$ evolution of the fragmentation functions. In this paper, we simply use
the FORTRAN code provided by Kretzer \cite{kcode} to get fragmentation
functions at any given $Q^2$. For each $Q^2$ we use Eqs. (\ref{2}-\ref{6}) to
fit the fragmentation functions and get the corresponding parameters for each
parton distributions at that scale $Q^2$.

In accordance with other theoretical results, we define here a dimensionless
scale $u=\ln (Q^2/\Lambda^2)$ with the QCD scale $\Lambda=0.19 {\rm GeV}/c$.
It is found that the parameters for each shower parton distribution has a
strong dependence on the scale $u$. As examples, the $Q^2$
dependence of parameters $A, a, b, c, d$ for $L$ and $K_{\rm NS}$,
obtained from fitting the fragmentation functions at different $Q^2$,
are shown in Fig. 3 as functions of $u=\ln (Q^2/\Lambda^2)$ by symbols.
For easy use of others, one can express the $Q^2$ dependence of
the parameters by a polynomial of the scale $u$ as
\begin{equation}
A(Q^2)=C_0u^3+C_1u^2+C_2u+C_3\ .
\label{q2}
\end{equation}
The fitted results are shown by solid curves also in Fig. 3 for comparison.
Then the dependence of every of five parameters for a shower parton distribution
in a jet is given by four parameters, $C_0, C_1, C_2$ and $C_3$ in a wide
range of $Q^2$ from 2 to 1000 $({\rm GeV}/c)^2$. Since the $Q^2$ dependence of
each parameter in the parton distributions can be well reproduced by only four
constant coefficients, Eq. (\ref{q2}) is a quite economic description of the evolution
of shower parton distributions. The coefficients $C_0, C_1, C_2$ and $C_3$ are tabulated
in Table I for all the parameters for the five parton shower distributions discussed
in this paper. From this table, parton distributions can be calculated easily
for a jet with any scale $Q^2$ in the range considered in this paper.

\section{Discussions}
The evolution of shower parton distributions in a jet is discussed
in the framework of quark recombination model from the $Q^2$ dependence
of the fragmentation functions. We show that the evolution of the distributions
can be given by the polynomial $\ln Q^2$ dependence of a few parameters.
From the tabulated coefficients one can calculate corresponding parton distributions
in a jet and other fragmentation functions needed in the framework of quark
recombination model. Together with soft and hard parton distributions, these
shower parton distributions can be used for the production of hadrons in
different collision processes.

\acknowledgments{This work was supported in part by the National Natural Science
Foundation of China under grant No. 10475032 and by the Ministry of Education of
China under grant No. 03113. C.B.Y would like to thank R.C. Hwa and X.N. Wang
for stimulating discussions.}

\begin{table}
\caption{Fitted coefficients in Eq. (\ref{q2}) for parameters of the parton
distributions in a jet.}
\begin{tabular}{l|c|c|c|c|c}
& & $C_0$ & $C_1$ & $C_2$ & $C_3$\\ \hline\hline
$K_{NS}$& $A$ & $2.27\times10^{-5}$ & $-4.91\times10^{-4}$ & $2.78\times10^{-3}$
& $5.92\times10^{-3}$\\ \hline
  &  $a$ &  $-7.21\times10^{-5}$ & $1.94\times10^{-3}$ &  $-1.92\times10^{-2}$
  & $8.46\times10^{-2}$\\ \hline
  &  $b$ &  $1.24\times10^{-3}$ &  $-3.47\times10^{-2}$ & $0.398$ &
  $-6.24\times10^{-3}$\\ \hline
  &  $c$ &  $-0.328$ & $7.67$ & $-45.3$ &  $246$\\ \hline
  &  $d$ &  $2.92\times10^{-4}$ & $-7.17\times10^{-3}$ &  $4.88\times10^{-2}$
  & $0.993$\\ \hline
$L$  & $A$ &   $-0.0009$ & $0.0275$ &  $-0.3179$ & $1.7997$\\ \hline
   & $a$ &   $-0.0003$ & $0.0116$ &  $-0.1489$ & $0.4376$\\ \hline
   & $b$ &   $0.0023$ &  $-0.0588$ & $0.5344$ &  $1.6756$\\ \hline
   & $c$ &   $0.0251$ &  $-0.5642$ & $3.8456$ &  $-2.7196$\\ \hline
   & $d$ &   $0.0040$ &  $-0.0773$ & $0.2938$ &  $5.2021$\\ \hline
$L_s$ &  $A$ &   $-0.0427$ & $0.9992$ &  $-7.7046$ & $21.3488$\\ \hline
   & $a$  & $-0.0049$ & $0.1210$ &  $-1.0107$ & $3.5884$\\ \hline
   & $b$ &  $-0.0022$ & $0.0442$ &  $-0.1268$ & $8.1004$\\ \hline
   & $c$ &   $0.1677$ &  $-4.2572$ & $38.9198$ & $-18.9039$\\ \hline
   & $d$ &   $-0.0014$ & $0.0308$ &  $-0.2036$ & $3.4815$\\ \hline
$G$ &  $A$ &   $-0.0025$ & $0.0835$ &  $-0.9312$ & $3.9350$\\ \hline
   & $a$ &  $0.0004$ &  $0.0007$ &  $-0.1318$ & $0.5742$\\ \hline
   & $b$ &  $-0.0015$ & $-0.0030$ & $0.4695$ &  $-0.2050$\\ \hline
   & $c$ &  $0.0003$ &  $-0.0073$ & $0.0575$ &  $-1.1224$\\ \hline
   & $d$ &  $0.0014$ &  $-0.0906$ & $1.0256$ &  $-1.6076$\\ \hline
$G_s$ & $A$ &   $-0.0015$ & $0.0319$ &  $-0.2183$ & $0.5004$\\ \hline
   & $a$ &  $-0.0140$ & $0.3064$ &  $-2.0830$ & $3.6414$\\ \hline
   & $b$ &  $-0.0061$ & $0.1095$ &   $-0.3339$ & $1.6614$\\ \hline
   & $c$ &  $0.1277$ &  $-2.7241$ & $18.0042$ & $-34.0906$\\ \hline
   & $d$ &   $-0.0128$ & $0.2812$ &  $-1.9713$ & $5.6142$\\ \hline\hline
\end{tabular}
\end{table}

\begin{thebibliography}{99}
\bibitem{frag} R.D. Field and R.P. Feynman, Nucl. Phys. {\bf B 136}, 1 (1978).

\bibitem{app} R.D. Field, {\it Applications of Perturbative QCD}, Addison-Wesley
Publishing Company (1989).

\bibitem{propion} S.S. Alder, PHENIX Collaboration, Phys. Rev. {\bf C 69}, 034909 (2004).

\bibitem{hy1} R.C. Hwa and C.B. Yang, Phys. Rev. {\bf C70}, 024904 (2004).

\bibitem{recom1} R.C. Hwa and C.B. Yang, Phys. Rev. {\bf C 67}, 034902 (2003);
V. Greco, C.M. Ko, and P. Levai, Phys. Rev. Lett. {\bf 90}, 202302 (2003);
Phys. Rev. {\bf C 68}, 024902 (2003); R.J. Fries, B. M\"uller, C. Nonaka,
and S.A. Bass, Phys. Rev. Lett. {\bf 90}, 202303(2003);
Phys. Rev. {\bf C 68}, 044902 (2003).

\bibitem{recom2} R.C. Hwa and C.B. Yang, Phys. Rev. Lett. {\bf 93}, 082302 (2004);
Phys. Rev. {\bf C 70}, 024905 (2004); Phys. Rev. {\bf C70}, 037901 (2004);
Phys. Rev. {\bf C 70}, 054902 (2004).

\bibitem{cronin} J.W. Cronin et al., Phys. Rev. {\bf D 11}, 3105 (1975).

\bibitem{rh}
R.C. Hwa, Phys. Rev. {\bf D 22}, 1593 (1980).

\bibitem{kkp}
B.A. Kniehl, G. Kramer and B. P\"otter, Nucl. Phys.
{\bf B 582}, 514 (2000).

\bibitem{kkp2}
B.A. Kniehl, G. Kramer and B. P\"otter, Nucl. Phys.
{\bf B 597}, 337 (2001).

\bibitem{sk}
S. Kretzer, Phys. Rev. {\bf D 62}, 054001 (2000).

\bibitem{valon} R.C. Hwa and C.B. Yang, Phys. Rev. {\bf C 66}, 025205 (2002).

\bibitem{dglap} Yu. Dokshitzer, Sov. Phys. JETP {\bf 46}, 1649 (1977); V.N. Gribov,
L.N. Lipatov, Sov. Nucl. Phys. {\bf 15}, 438 (1972); V.N. Gribov,
L.N. Lipatov, Sov. Nucl. Phys. {\bf 15}, 675 (1972); G. Altarelli and G. Parisi, Nucl.
Phys. {\bf B 126}, 298 (1977).

\bibitem{kcode} See the website http://www.pv.infn.it/$\sim$radici/FFdatabase/

\end{thebibliography}
\end{document}